\documentclass[reprint,onecolumn,notitlepage,amsmath,aps,nofootinbib]{revtex4-1}
\usepackage{graphicx}
\usepackage{dcolumn}
\usepackage{bm}
\usepackage{multirow}
\begin{document}

\title{CMB-S4 forecast on the primordial non-Gaussianity parameter of feature models}

\author{Wuhyun Sohn}
\email{ws313@damtp.cam.ac.uk}
\author{James Fergusson}
\affiliation{
	Centre for Theoretical Cosmology, DAMTP, University of Cambridge, Cambridge CB3 0WA, United Kingdom
}
\date{\today}

\begin{abstract}
	
	We present forecasts on the primordial non-Gaussianity parameter $f_\text{NL}$ of feature models for the future Cosmic Microwave Background Stage-4 (CMB-S4) experiments. The Fisher matrix of the bispectrum estimator was computed using noise covariances expected for preliminary CMB-S4 specifications including ones for the Simons Observatory. We introduce a novel method that improves the computation by orthonormalising the covariance matrix. The most sensitive CMB-S4 experiment with 1' beam and 1$\mu K$-arcmin noise would yield a factor of 1.7-2.2 times more stringent constraints compared to Planck. Under the Simons Observatory baseline conditions the improvement would be about 1.3-1.6 times to Planck. We also thoroughly studied the effects of various model and experimental parameters on the forecast. Detailed analysis on the constraints coming from temperature and E-mode polarisation, in particular, provided some insight into detecting oscillatory features in the CMB bispectrum.
	
\end{abstract}

\maketitle

\section{Introduction}

The Cosmic Microwave Background (CMB) radiation is one of our most valuable probes of the primordial universe. The temperature and polarisation of this ancient light contains rich statistical information both about the primordial perturbations created during inflation and also their subsequent evolution until now. This allows us to test our inflationary theories and also the history of our universe. The recent Planck CMB experiments have provided stringent tests on various models of inflation through the estimation of cosmological parameters and via primordial non-Gaussianity \cite{PlanckCollaboration2015,PlanckCollaboration2018}.

The simplest model of inflation involves a single scalar field slowly rolling down a smooth potential. In this case the CMB temperature fluctuations are expected to be Gaussian distributed with only tiny deviations (e.g. \cite{Maldacena2013}). However, many other physically well-motivated models generate larger non-Gaussian signatures at the end of inflation (see reviews of \cite{Chen2010}). Such primordial non-Gaussianities are well constrained by three-point correlation functions of the CMB anisotropies or their Fourier transform, the CMB bispectrum. Different inflationary models predict bispectra with different momentum dependence, or `shapes'. We constrain these models by using an optimal estimator for their amplitude parameter,  $f_\text{NL}$,  for each specific bispectrum shape (see, e.g., \cite{Komatsu2010,Liguori2010} for reviews).

Although all observations to date are consistent with vanishing non-Gaussianity, the models most favoured by the 2015 Planck CMB analysis were the ones with oscillations in the primordial power spectrum \cite{PlanckCollaboration2015}. Among them are feature models, where the oscillations are caused by a sharp feature in either the inflationary potential \cite{starobinsky1992,Adams2001,Chen2007,Adshead2012,Hazra2014,Dvorkin2010}, sound speed \cite{Miranda2012,Bartolo2013}, or multi-field potentials \cite{Achucarro2011} (see \cite{Chen2010,Chluba2015} for reviews). The primordial power spectrum then becomes scale dependent, displaying sinusoidal oscillations that are linearly spaced in momentum space. The resulting bispectrum also oscillates and is highly uncorrelated with other popular bispectrum templates \cite{Meerburg2009}, therefore allowing us to constrain them independently.

Planck constrained $f_\text{NL}$ for feature models from CMB bispectra, but no signal above $3\sigma$ significance were found after accounting for the `look elsewhere effect' as introduced in \cite{Fergusson2015a}. The multi-peak statistic analysis, however, revealed some non-standard signals up to $4\sigma$ level that deserves attention \cite{PlanckCollaboration2015}. There have been many other searches on signatures of oscillations. Constraints also come from the CMB power spectrum \cite{Martin2004,Benetti2011,Meerburg2012,Meerburg2014a,Meerburg2014,Fergusson2015b}, the large scale structure \cite{Chantavat2011,Ballardini2016}, and a combination of the two \cite{Hu2015,Benetti2016}. We expect stronger constraints on feature models from future LSS experiments \cite{Chen2016}. This paper covers the prospects of upcoming CMB experiments in constraining $f_\text{NL}$ for feature models.

Currently there are two implementations of the optimal estimator for constraining $f_\text{NL}$ for feature type models. The Planck analysis adopted the Modal estimator for which the given bispectrum is expanded using a separable basis \cite{Fergusson2012,Fergusson2014}. This method is efficient, can flexibly account for various oscillatory shapes, and is able to easily constrain all frequencies simultaneously. However, when the oscillation frequency is large the modal basis fails to converge within reasonable number of basis elements, making the method impractical. The other approach using the Komatsu-Spergel-Wandelt (KSW) estimator is viable for various shapes including the feature model \cite{Komatsu2005,Munchmeyer2014}. Although this method only applies to models with separable bispectra, even highly oscillatory templates can be computed reliably. This method is however more computationally expensive as each frequency must be dealt with separately. We present further optimisations to the fast KSW estimator introduced in \cite{Yadav2007} and apply it on feature models for forecasts in this paper.

The next generation of CMB experiments, CMB Stage-4, consists of many exciting proposed experiments located at the South Pole, the Atacama Desert in Chile, and perhaps space \cite{Abazajian2016,TheSimonsObservatoryCollaboration2018,TheCOrEcollaboration2015}. One of the main goals of these experiments is to measure the polarisation signal in the CMB to the cosmic variance limit. Preliminary specifications have been released for these experiments \cite{Abazajian2016,TheSimonsObservatoryCollaboration2018} and these have been used to produce some forecasts for the standard $f_\text{NL}$ templates but not yet for feature type models. In this paper we address this by presenting the Fisher forecasts on $f_{NL}$ for feature models based on these specifications and observe that feature type models receive larger improvements from the extra polarisation information than the standard templates, justifying this analysis.

The paper is organised as follows. First we briefly review the theory of CMB bispectrum in Section \ref{section: feature model bispectrum}. Bispectrum template for the feature model is defined and computed here. In Section \ref{section: Efficient computation of KSW estimator with polarisation} we formulate the bispectrum estimator and introduce a new method to further optimise its computation. The technique is applied to the case of feature model to yield equations for the Fisher forecast of $f_{NL}$. We also briefly discuss implementation details. In Section \ref{section: CMB-S4 forecast results} we present our forecast results and their dependence on model and experimental parameters. In particular, forecasts for the Simons observatory are compared with the Planck results. The results are summarised in Section \ref{section: conclusion}.

\section{Feature model bispectrum} \label{section: feature model bispectrum}
\subsection*{CMB bispectrum}

One of the main subjects of primordial non-Gaussianity studies is the 3-point correlation function of the primordial perturbations which is defined by;
\begin{equation}
\left< \Phi(\mathbf{k}_1) \Phi(\mathbf{k}_2) \Phi(\mathbf{k}_3) \right> = (2\pi)^3 \delta^{(3)}(\mathbf{k}_1 + \mathbf{k}_2 + \mathbf{k}_3) B_\Phi (k_1, k_2, k_3),
\end{equation}
where we have assumed statistical homogeneity and isotropy. The primordial bispectrum $B_\Phi$ vanishes for Gaussian perturbations, but more general inflation models predict non-zero bispectra with various shapes.  In order to constrain these models we re-parameterise the bispectrum into a amplitude parameter and a normalised shape part;
\begin{equation}
	B_\Phi (k_1, k_2, k_3) = f_{NL} B_\Phi^{(f_{NL}=1)} (k_1, k_2, k_3).
	\label{fNL definition}
\end{equation}
Constraining $f_{NL}$ from the CMB measurements allows to determine how well the particular shape under consideration aligns with the data, which we can then translate into constraints on the model itself.

In order to compare the theory with measurements we first need to relate the primordial perturbations to spherical multipole modes of the late-time CMB anisotropies.
\begin{equation}
a_{lm}^{X} = 4\pi(-i)^l \int \frac{d^3\mathbf{k}}{(2\pi)^3}  \Phi(\mathbf{k}) \Delta_l^X(k) Y_{lm}(\hat{\mathbf{k}}).
\label{theoretical multiple moments}
\end{equation}
Here the index $X$ is either $T$ or $E$, representing CMB temperature and E-mode polarisation, respectively. The linear CMB radiation transfer function $\Delta_l^X(k)$ can be computed from the Boltzmann solvers like CAMB \cite{Lewis2000}.

Three point correlation function of $a_{lm}^X$'s yield the reduced bispectrum $b_{l_1 l_2 l_3}$ times a geometrical factor $\mathcal{G}^{l_1 l_2 l_3}_{m_1 m_2 m_3}$ named the Gaunt integral. After some algebraic manipulations we obtain the following useful formula for the reduced bispectrum;
\begin{equation}
b_{l_1 l_2 l_3}^{X_1 X_2 X_3} = \left(\frac{2}{\pi}\right)^3 \int_{0}^{\infty} r^2 dr \int_{\mathcal{V}_k} d^3\textbf{k} \; (k_1 k_2 k_3)^2 B_\Phi (k_1, k_2, k_3) \prod_{i=1}^{3} \left[ j_{l_i}(k_i r) \, \Delta_{l_i}^{X_i}(k_i) \right],
\label{reduced bispectrum}
\end{equation}
where $j_l$ is the spherical Bessel function arising from the Rayleigh expansion formula. Using this equation, we can compute the projected bispectrum from any given primordial bispectrum. Direct computation of this four-dimensional integral for every $l$ combination, however, is practically impossible. Not only is the integral in 4D but also the oscillatory integrand requires a large number of sample points in each of $k_i$, making the full calculation for every $l_i$ triple prohibitively expensive. All bispectrum estimators get around this problem by expanding $B_\Phi$ as a sum of \textit{separable} terms. This will be explained in more detail later using the feature model template as an example.

\subsection*{Feature model}

We follow the works of \cite{Munchmeyer2014,Fergusson2015a,Fergusson2015b,PlanckCollaboration2015} and assume the following template for the bispectrum of feature models;
\begin{equation}
B_\Phi ^ {\text{feat}}(k_1, k_2, k_3) = \frac{6 A^2}{(k_1 k_2 k_3)^2} \sin\,(\omega K + \phi),
\label{feature model definition}
\end{equation}
where $K = k_1 + k_2 + k_3$, $A$ represents the primordial power spectrum amplitude, and $\phi$ is a phase. The oscillation `frequency' $\omega$ is associated with the location and scale of feature in the inflationary potential. It is often written in terms of the oscillation scale $k_c$ as $\omega = 2\pi/3k_c$. $\omega$ is measured in Mpc but we omit the unit for notational conveniences.

The feature model template has two free parameters that need to be fixed before we can constrain the model: $\omega$ and $\phi$. The phase $\phi$ can be easily dealt with by observing that
\begin{equation}
	B_\Phi^\text{feat}(k_1, k_2, k_3) = \cos\phi \; B_\Phi^\text{sin} (k_1, k_2, k_3) + \sin\phi \; B_\Phi^\text{cos} (k_1, k_2, k_3).
	\label{feature model bispectrum as a sum of sin and cos}
\end{equation}
Here $B_\Phi^\text{sin}$ and $B_\Phi^\text{cos}$ correspond to feature models with $\phi = 0$ and $\pi/2$, respectively. Non-zero phase simply corresponds to a linear combination of the sine and cosine templates. As we will see later these two shapes are in fact highly uncorrelated. Therefore, they can be constrained independently from each other.

On the other hand, one still has a complete freedom of choice on the oscillation frequency $\omega$. Such freedom dramatically expands size of the parameter space. In practice we constrain $f_\text{NL}$ for each fixed value of oscillation frequency, which yields hundreds of estimates. Since there are so many estimates we are looking at, there is a good chance that we find notable signals by sheer luck. Accounting for this `look elsewhere effect' has been resolved using methods in \cite{Fergusson2015a} and subsequently applied to the Planck analysis \cite{PlanckCollaboration2015,Fergusson2015b}. The look-elsewhere-adjusted statistics used in the literature can be employed for the future CMB-S4 data analysis. This work, however, focusses on forecasting the `raw' estimates and comparing them with those of Planck.

\subsection*{Separability}

The bispectrum template of feature models (\ref{feature model definition}) is an example of separable shape. It can be expressed as a sum of terms in the form $f(k_1)g(k_2)h(k_3)$ for some functions $f$, $g$ and $h$, which dramatically simplifies the computation of reduced bispectrum $b_{l_1 l_2 l_3}$. The three-dimensional integral over the $k$ space in (\ref{reduced bispectrum}) splits into three individual one-dimensional integrals for separable shapes. Feature models for example has
\begin{eqnarray}
b_{l_1 l_2 l_3}^{X_1 X_2 X_3,\text{feat}} &=& 6A^2 \left( \frac{2}{\pi} \right)^3 \int_0^\infty r^2 dr \int_{\mathcal{V}_k} d^3\mathbf{k}\, e^{i\omega (k_1 + k_2 + k_3)} \prod_{i=1}^{3} \left[ j_{l_i} (k_i r) \Delta_{l_i}^{X_i} (k_i) \right] \nonumber \\
&=&  6A^2 \left( \frac{2}{\pi} \right)^3 \int_0^\infty r^2 dr \, \prod_{i=1}^{3} \left[ \int_0^\infty dk_i \, e^{i \omega k_i} j_{l_i} (k_i r) \Delta_{l_i}^{X_i} (k_i) \right]. \label{feature model reduced bispectrum}
\end{eqnarray}
Here the real and imaginary parts of $b^\text{feat}$ correspond to the bispectra of cosine and sine feature models, respectively. Now define
\begin{eqnarray}
s_l^X(r) &:=& \frac{2A^{2/3}}{\pi} \int_0^\infty dk \, \sin\,(\omega k) j_l(kr) \Delta_l^X(k) \\
c_l^X(r) &:=& \frac{2A^{2/3}}{\pi} \int_0^\infty dk \, \cos\,(\omega k) j_l(kr) \Delta_l^X(k).
\end{eqnarray}
These are analogous to $\alpha_l^X(r)$ and $\beta_l^X(r)$ in the usual KSW estimator for local non-Gaussianity. Then (\ref{feature model reduced bispectrum}) reduces to
\begin{eqnarray}
b_{l_1 l_2 l_3}^{X_1 X_2 X_3, \text{feat}} &=& 6 \int_0^\infty r^2 dr \, \left(c_{l_1}^{X_1} c_{l_2}^{X_2} c_{l_3}^{X_3} -  c_{l_1}^{X_1} s_{l_2}^{X_2} s_{l_3}^{X_3} -  s_{l_1}^{X_1} c_{l_2}^{X_2} s_{l_3}^{X_3} -  s_{l_1}^{X_1} s_{l_2}^{X_2} c_{l_3}^{X_3} \right)  \nonumber \\
&\quad& +\; 6i  \int_0^\infty r^2 dr \, \left(s_{l_1}^{X_1}c_{l_2}^{X_2}c_{l_3}^{X_3} + c_{l_1}^{X_1}s_{l_2}^{X_2}c_{l_3}^{X_3} +  c_{l_1}^{X_1}c_{l_2}^{X_2}s_{l_3}^{X_3} -  s_{l_1}^{X_1}s_{l_2}^{X_2}s_{l_3}^{X_3} \right). 	\nonumber \\
&=& b_{l_1 l_2 l_3}^{X_1 X_2 X_3, \text{cos}} + i \; b_{l_1 l_2 l_3}^{X_1 X_2 X_3, \text{sin}}
\label{feature model reduced bispectrum sin and cos}
\end{eqnarray}

\section{Efficient computation of the estimator with polarisation} \label{section: Efficient computation of KSW estimator with polarisation}

\subsection*{Estimator}

The optimal estimator for a given bispectrum in the weak non-Gaussian limit involves computing \cite{Komatsu2005,Komatsu2010}
\begin{eqnarray}
S_i =  \frac{1}{6} \sum_{l_j, m_j} \sum_{X_j} \mathcal{G}_{m_1 m_2 m_3}^{l_1 l_2 l_3} b_{l_1 l_2 l_3} ^{X_1 X_2 X_3, (i)} (C_{l_1 m_1, l_4 m_4}^{-1})^{X_1 X_4} (C_{l_2 m_2, l_5 m_5}^{-1})^{X_2 X_5} (C_{l_3 m_3, l_6 m_6}^{-1})^{X_3 X_6} \nonumber \\  \left[ a_{l_4 m_4}^{X_4} a_{l_5 m_5}^{X_5} a_{l_6 m_6}^{X_6} -  \left( C_{l_4 m_4, l_5 m_5} a_{l_6 m_6}^{X_6} + \text{2 cyclic} \right)   \right].
\label{Estimator full definition}
\end{eqnarray}
Here summations are over $l_j$, $m_j$, $X_j$ and $X'_j$ for each $j=1,2,3$. The spherical multipole moments $a_{lm}^{X}$'s are computed from observations, and $b^{(i)}$ denotes the $i$th theoretical bispectrum template under consideration.

Computing this form, however, requires an inversion of the full covariance matrix $C_{lm,l'm'}$, which is computationally expensive. As a result we will follow the diagonal covariance approximation in \cite{Yadav2007} for the inverse covariance; $C_{l_1 l_4 m_1 m_4}^{-1} \approx (1/C_{l_1})~ \delta^D_{l_1 l_4} \delta^D_{m_1 -m_4}$. We also approximate the covariance in the linear term by an ensemble average over Monte Carlo simulations of Gaussian realisations; $C_{l_4 l_5 m_4 m_5}^{X_1 X_2} \approx \left< a_{l_1 m_1}^{X_1} a_{l_2 m_2}^{X_2} \right>$.  With these simplifications the estimator takes the form
\begin{equation}
\hat{f_i} = \sum_j (F^{-1})_{ij} S_j,
\label{Estimator f definition}
\end{equation}
where
\begin{eqnarray}
S_i =  \frac{1}{6} \sum_{l_j, m_j} \sum_{X_j, X_j'} \mathcal{G}_{m_1 m_2 m_3}^{l_1 l_2 l_3} b_{l_1 l_2 l_3} ^{X_1 X_2 X_3, (i)} (C_{l_1}^{-1})^{X_1 X_1'} (C_{l_2}^{-1})^{X_2 X_2'} (C_{l_3}^{-1})^{X_3 X_3'} \nonumber \\  \left[ a_{l_1 m_1}^{X_1'} a_{l_2 m_2}^{X_2'} a_{l_3 m_3}^{X_3'} -  \left( \left< a_{l_1 m_1}^{X_1'} a_{l_2 m_2}^{X_2'} \right> a_{l_3 m_3}^{X_3'} + \text{2 cyclic} \right)   \right],
\label{Estimator S definition}
\end{eqnarray}
and
\begin{equation}
F_{ij}= \frac{f_\text{sky}}{6} \sum_{\text{all }X, X'} \sum_{\text{all }l} h_{l_1 l_2 l_3}^2 \; b_{l_1 l_2 l_3}^{X_1 X_2 X_3, (i)} (C_{l_1}^{-1})^{X_1 X_1'} (C_{l_2}^{-1})^{X_2 X_2'} (C_{l_3}^{-1})^{X_3 X_3'} \; b_{l_1 l_2 l_3}^{X_1' X_2' X_3', (j)}.
\label{Estimator F definition}
\end{equation}
The covariance matrix $C_l$ is now a $2\times2$ matrix consisting of values $C_l^{TT}$, $C_l^{TE}$, $C_l^{ET}$ and $C_l^{EE}$. \footnote{Note that this is equivalent to having a $2l\times2l$ matrix with diagonal $l\times l$ block matrices $C^{TT}$, $C^{TE}$, $C^{ET}$ and $C^{EE}$ as in other literatures including \cite{Fergusson2014}.} The linear terms (the second in square brackets) are required to account for anisotropies induced by masking and anisotropic noise.

$F_{ij}$ is the Fisher information matrix of the estimator. $f_\text{sky}$ in (\ref{Estimator F definition}) denotes the fraction of the sky covered by the experiment, and $h_{l_1 l_2 l_3}^2 := \sum_{m_j} \left( \mathcal{G}_{m_1 m_2 m_3}^{l_1 l_2 l_3} \right){}^2$ is a geometric factor. Since the estimator $\hat{f}_i$ in (\ref{Estimator f definition}) is nearly optimal, its 68\% confidence (1$\sigma$) interval can be computed from the Fisher matrix as $\sigma_i := \Delta f_{NL}^{(i)} = (F^{-1})_{ii}$.

Note that most CMB-S4 experiments are ground-based, so they can probe smaller fraction of the sky compared to Planck. Having a smaller fraction of the sky leads to  increased uncertainties for the estimator. Current estimate is that the new experiments will cover 40\% of the sky, significantly less than the 74\% of Planck. The error bars will thus increase by a factor of 1.38 from the decrease in $f_\text{sky}$ alone.  This may be reduced by combining Planck data for unobserved pixels in these experiments

\subsection*{Orthonormalising the covariance matrix}

In \cite{Fergusson2014} it was noted that orthogonalising the multipoles of temperature and polarisation maps dramatically reduces the number of terms in computation of the Modal estimators. This technique can also be applied to KSW estimators, or indeed any optimal bispectrum estimator, which is yet to be done to the authors' knowledge.

In both (\ref{Estimator S definition}) and (\ref{Estimator F definition}) there are summations over indices $X$ and $X'$ to account for correlations between the CMB temperature and E-mode polarisation. This can be simplified by essentially making a change of basis in $X$ space for each $l$ so that every $C_l$ becomes orthonormal. We perform a Cholesky decomposition on $C_l$ and invert the matrix. Then $C_l^{-1} = L_l^T L_l$, where $L_l$ is a lower triangular matrix given by
\begin{equation}
	L_l = \begin{pmatrix} \frac{1}{\sqrt{C_l^{TT}}} & 0  \\ \frac{- C_l^{TE}}{\sqrt{C_l^{TT}} \sqrt{C_l^{TT} C_l^{EE} - \left( C_l^{TE} \right)^2 }}  &  \frac{ C_l^{TT}}{\sqrt{C_l^{TT}} \sqrt{C_l^{TT} C_l^{EE} - \left( C_l^{TE} \right)^2}} \end{pmatrix}
\end{equation}
Now let
\begin{equation}
	\tilde{\Delta}_{l}^X (k) = \sum_{X'} L_l^{XX'} \Delta_{l}^{X'} (k), \quad\text{and}\quad \tilde{a}_{lm}^X = \sum_{X'} L_l^{XX'} a_{lm}^{X'}.
	\label{orthonormalisation}
\end{equation}

Defining $\tilde{b}_{l_1 l_2 l_3}$ to be the corresponding reduced bispectrum, (\ref{Estimator S definition}) and (\ref{Estimator F definition}) simplify to
\begin{eqnarray}
S_i =  \frac{1}{6} \sum_{l_j, m_j} \sum_{X_j} \mathcal{G}_{m_1 m_2 m_3}^{l_1 l_2 l_3} \tilde{b}_{l_1 l_2 l_3} ^{X_1 X_2 X_3, (i)} \left[ \tilde{a}_{l_1 m_1}^{X_1} \tilde{a}_{l_2 m_2}^{X_2} \tilde{a}_{l_3 m_3}^{X_3} -  \left( \left< \tilde{a}_{l_1 m_1}^{X_1} \tilde{a}_{l_2 m_2}^{X_2} \right> \tilde{a}_{l_3 m_3}^{X_3} + \text{2 cyclic} \right)   \right],
\label{KSW estimator S after orthonormalisation}
\end{eqnarray}
\begin{equation}
F_{ij}= \frac{f_\text{sky}}{6} \sum_{\text{all }X} \sum_{\text{all }l} h_{l_1 l_2 l_3}^2 \; \tilde{b}_{l_1 l_2 l_3}^{X_1 X_2 X_3, (i)} \; \tilde{b}_{l_1 l_2 l_3}^{X_1 X_2 X_3, (j)}.
\label{KSW estimator F after orthonormalisation}
\end{equation}

Using this method not only makes it more mathematically concise, but also halves the number of terms involved in the summation. Linear transformations (\ref{orthonormalisation}) only need to be done once in the beginning of the program and cost little compared to the main computation. We also found it easier to optimise the code using instruction level vectorisations after this simplification.

The only downside of this method is that we no longer can get breakdowns of signal from each of $TTT$, $TTE$, $TEE$ and $EEE$ bispectrum since our new modes are linear combinations of $T$ and $E$ modes. However, in most cases we are interested in either $T$-only or $T+E$ results, and this method works perfectly well in these cases.

\subsection*{Estimator for feature models}

We compute the general estimator (\ref{KSW estimator S after orthonormalisation}) and (\ref{KSW estimator F after orthonormalisation}) for feature models. The method is similar to the one seen in \cite{Munchmeyer2014} except that now the polarisation is included and the covariance matrices are trivial thanks to the orthonormalisation process outlined above.

Consider the bispectrum shape of
\begin{equation}
	B_\Phi (k_1, k_2, k_3) = f_{NL}^{\sin} B^{\sin}(k_1, k_2, k_3) + f_{NL}^{\cos} B^{\cos}(k_1, k_2, k_3),
\end{equation}
for a fixed value of oscillation frequency $\omega$. Here $B^{\sin}$ and $B^{\cos}$ correspond to reduced bispectra $b^{\sin}$ and $b^{\cos}$ defined in (\ref{feature model reduced bispectrum sin and cos}). The Fisher matrix $F$ is $2\times2$ but its off-diagonal entries are 2-3 orders of magnitude smaller than diagonal ones in most cases as will be presented in the next section. Thus, the two shapes are assumed to be uncorrelated and constrained individually. Here we present detailed computations for $f_{NL}^{\sin}$ only but the cosine one can be computed similarly.

From (\ref{feature model reduced bispectrum sin and cos}) and the definition of Gaunt integral $\mathcal{G}^{l_1 l_2 l_3}_{m_1 m_2 m_3} = \int d\hat{\mathbf{n}} \; Y_{l_1 m_1}(\hat{\mathbf{n}}) Y_{l_2 m_2}(\hat{\mathbf{n}}) Y_{l_3 m_3}(\hat{\mathbf{n}})$ it follows that

\begin{equation}
S^\text{cub} = \int_0^\infty r^2 dr \int d^2\hat{\mathbf{n}} \left[ - {M_s}^3 + 3 M_s {M_c}^2 \right] \quad\text{and}
\end{equation}
\begin{equation}
S^\text{lin} = -3 \int_0^\infty r^2 dr \int d^2\hat{\mathbf{n}} \left[ - {M_s} \left<{M_s}^2\right> + M_s \left< {M_c}^2 \right> + 2 {M_c} \left< {M_s} {M_c} \right> \right],
\end{equation}
where
\begin{eqnarray}
M_s(r, \hat{\mathbf{n}}) &:=& \sum_X \sum_{l m} \tilde{s}_l^X(r) \, \tilde{a}_{lm}^X \, Y_{lm}(\hat{\mathbf{n}}), \nonumber\\
M_c(r, \hat{\mathbf{n}}) &:=& \sum_X \sum_{l m} \tilde{c}_l^X(r) \, \tilde{a}_{lm}^X \, Y_{lm}(\hat{\mathbf{n}}).
\label{M_s definition}
\end{eqnarray}
Again, the bracket $\left< \cdot \right>$ denotes averaging over Gaussian simulations. The sum of $S^\text{cub}$ and $S^\text{lin}$ gives the final value of $S$ for sine feature model.

For efficient Fisher matrix calculation we follow \cite{Smith2011} and deploy the identity
\begin{equation}
h_{l_1 l_2 l_3} ^2 = \frac{(2l_1 +1)(2l_2 +1)(2l_3 +1)}{8\pi} \int_{-1}^{1} d\mu \, P_{l_1}(\mu)P_{l_2}(\mu)P_{l_3}(\mu), \label{h}
\end{equation}
where $P_{l}(\mu)$ represents the Legendre polynomial. Then,
\begin{equation}
F = \frac{3}{4\pi} \int r^2 dr \int r'^2 dr' \int d\mu \left[ P_{ss}^3 + 3 P_{ss} P_{cc}^2 - 3 P_{cs}^2 P_{ss} - 3 P_{sc}^2 P_{ss} + 6 P_{cs}P_{sc}P_{cc} \right].
\end{equation}
where we have defined
\begin{eqnarray}
P_{ss}(r, r', \mu) &:=& \sum_{X} \sum_{l} (2l+1) \, \tilde{s}_l^X(r) \tilde{s}_l^X(r') P_l(\mu) \nonumber \\
P_{sc}(r, r', \mu) &:=& \sum_{X} \sum_{l} (2l+1) \, \tilde{s}_l^X(r) \tilde{c}_l^X(r') P_l(\mu).
\label{P_ss definition}
\end{eqnarray}
and similarly $P_{cs}$ and $P_{cc}$.

Calculations of (\ref{M_s definition}) and (\ref{P_ss definition}) are two of the most computationally expensive steps. If we have not orthonormalised the covariance matrix, then there would be an extra summation over $X'$ and some $2\times2$ matrix algebra involving $(C_l^{-1})^{XX'}$.

\subsection*{Probing beam and instrumental noise}

In an ideal experiment where measurements are made on each point of the sky perfectly, the covariance matrice $C_l^{XX'}$ in (\ref{Estimator S definition}) and (\ref{Estimator F definition}) consists purely of the signal. In reality, however, the probing beam has finite width and the sensors are noisy. These effects can be incorporated by modifying the covariance matrices and bispectra as follows.
\begin{equation}
C_l^{X_1 X_2} \rightarrow\; W_l^{X_1} W_l^{X_2} C_l^{X_1 X_2} + N_l^{X_1 X_2} ,\qquad b_{l_1 l_2 l_3}^{X_1 X_2 X_3} \rightarrow\; W_{l_1}^{X_1} W_{l_2}^{X_2} W_{l_3}^{X_3} b_{l_1 l_2 l_3}^{X_1 X_2 X_3},
\end{equation}
where $W_l^X$ and $N_l^{X_1 X_2}$ represent the beam window function and the noise covariance matrix, respectively. When substituted into the KSW estimator, these changes are equivalent to modifying
\begin{eqnarray}
C_l^{X_1 X_2} &\rightarrow& C_l^{X_1 X_2} + \left( W_l^{X_1} W_l^{X_2} \right)^{-1} N_l^{X_1 X_2} \nonumber \\ &=& ( C_l^\text{sig} )^{X_1 X_2} + ( C_l^\text{noise} )^{X_1 X_2},
\end{eqnarray}
while keeping the bispectra same. Here we have defined the effective (beam-corrected) noise covariance matrix $C_l^\text{noise}$. Modes for which $C_l^\text{noise}$ is much larger than $C_l^\text{sig}$ contribute little to the $f_{NL}$ estimator.

For forecasting purposes we assume Gaussian beam and white uncorrelated noise until more detailed experiment specifications become available. Under these assumptions, the effective noise covariances reduce to \cite{Ng1999}
\begin{eqnarray}
C_l^{\text{noise}, TT} = \exp\left({l(l+1)\sigma_\text{beam}^2} \right)N_\text{white}, \quad C_l^{\text{noise}, EE} =2\; C_l^{\text{noise}, TT}, \quad C_l^{\text{noise}, TE} = 0.
\end{eqnarray}
The factor of two for $EE$ mode is comes from measuring two Stokes parameters Q and U. The Gaussian beam profile is usually specified by its FWHM (full width at half maximum) in $arcmin$, which is then converted to standard deviations in radians for $\sigma_\text{beam}$. The noise level often comes in the units of $\mu K\cdot arcmin$. This is then divided by $T_\text{CMB} = 2.725K$, converted to radians and squared to get $N_\text{white}$.

For the Planck experiment, using 5 arcmin FWHM beam and the ~47 $\mu K \cdot $arcmin noise level gives good approximations to the post-component-separation noise covariances. For CMB-S4 experiments the details are not confirmed, but the beam FWHM is expected to lie between 1-5 arcmin, while the noise level will range from 1 to 9 $\mu K \cdot $arcmin. \cite{Abazajian2016}

In real measurements there exist extra contaminations in large angular scales due to 1/f noises and the component separation process. Though most of our analysis assumes simpler form of noise covariances elaborated above, for the Simons Observatory forecasts we follow \cite{TheSimonsObservatoryCollaboration2018} and model 1/f noise as $N_l = N_{red} (l/l_{knee})^{\alpha_{knee}} + N_{white}$. The noise curves from each channel were then put together using the inverse variance method. This is a good approximation for the E mode polarisation but not for temperature, since extra degradations occur during the component separation process. Still, because dominant contributions to the feature model signal comes from polarisation data, this would be a reasonable approximation for our forecast. For Planck the full post-component-separation noise curves are available and hence used for computations.

\subsection*{Implementation and validation}

We implemented the pipeline outlined above using the C programming language and parallelised using hybrid MPI + openMP. The code was then run in the COSMOS supercomputing system.

The transfer functions are generated from the CAMB code \cite{Lewis2000}. Bessel function values were pre-computed using recursion relations and stored in a file, while the Legendre function values were computed on the fly using the GNU scientific library. The angular power spectrum data was generated from $\Lambda$CDM parameters estimated in the Planck 2015 results. 

Numerical integration for variables $k$, $r$ and $r'$ were done using simple trapezoidal methods, as they can be easily vectorised for optimisation. On the other hand, integration of $\mu$ required more care because the Legendre polynomials are highly oscillatory. We adopted the Gauss-Legendre quadrature rule with $1.5 \,l_\text{max}+1$ points which can integrate polynomials up to order $3\,l_\text{max}$ exactly. The weights and nodes were computed in the beginning using the QUADPTS code \cite{Hale2013}.

Various checks have been done to ensure that the code runs correctly. First we used the code to reproduce the Planck results, which agreed within 3\% error. The code was then used to compute bispectrum for the constant model, corresponding to the case where $\omega=\phi=0$. There exists an approximate analytic form in this case \cite{Fergusson2012} which we were able to reproduce accurately. We also performed convergence tests on $r$ and $r'$ integration by doubling the number of points for each of them. The grid was chosen to be very dense around recombination and quite dense near reionisation. We confirmed that changes in the integral are less than 0.5\% for each value of $\omega$.

\section{CMB-S4 forecast results} \label{section: CMB-S4 forecast results}
\subsection*{Phase dependence}

We now present the CMB-S4 forecast on the error bars of primordial non-Gaussianity parameter for feature models. For notational convenience we denote the error bars for sine and cosine feature models by $\sigma_{\sin}$ and $\sigma_{\cos}$. Superscripts $T$ and $T+E$ are also put to distinguish temperature-only analysis from the full analysis including polarisation.

First of all, we check that the sine and cosine bispectrum templates defined in (\ref{feature model definition}) are indeed uncorrelated and can be constrained separately. In order to do this, we see if the Fisher matrix of feature models is robust to changes in the phase for different $\omega$ values of interest. Feature model bispectra with a specific phase $\phi$ can be represented as a sum of sine and cosine ones as in (\ref{feature model bispectrum as a sum of sin and cos}). Hence, its Fisher matrix is given by
\begin{equation}
	F(\omega, \phi) = \cos^2\phi \; F_{ss}(\omega) + \sin^2\phi \; F_{cc}(\omega) + 2\cos\phi\sin\phi \; F_{sc}(\omega),
\end{equation}
where $F_{ss}$ is the element $F_{ij}$ of the Fisher matrix in (\ref{Estimator F definition}) with reduced bispectra $b^{(i)} = b^{(j)} = b^{\sin}$, and so on. Correlation between sine and cosine templates can be expressed as $F_{sc}/(F_{ss}F_{cc})^{1/2}$, and this value can be learned from analysing the $\phi$ dependence of $F(\omega,\phi)$.

\begin{figure}[ht]
	\centering
	\includegraphics[width=0.5\textwidth]{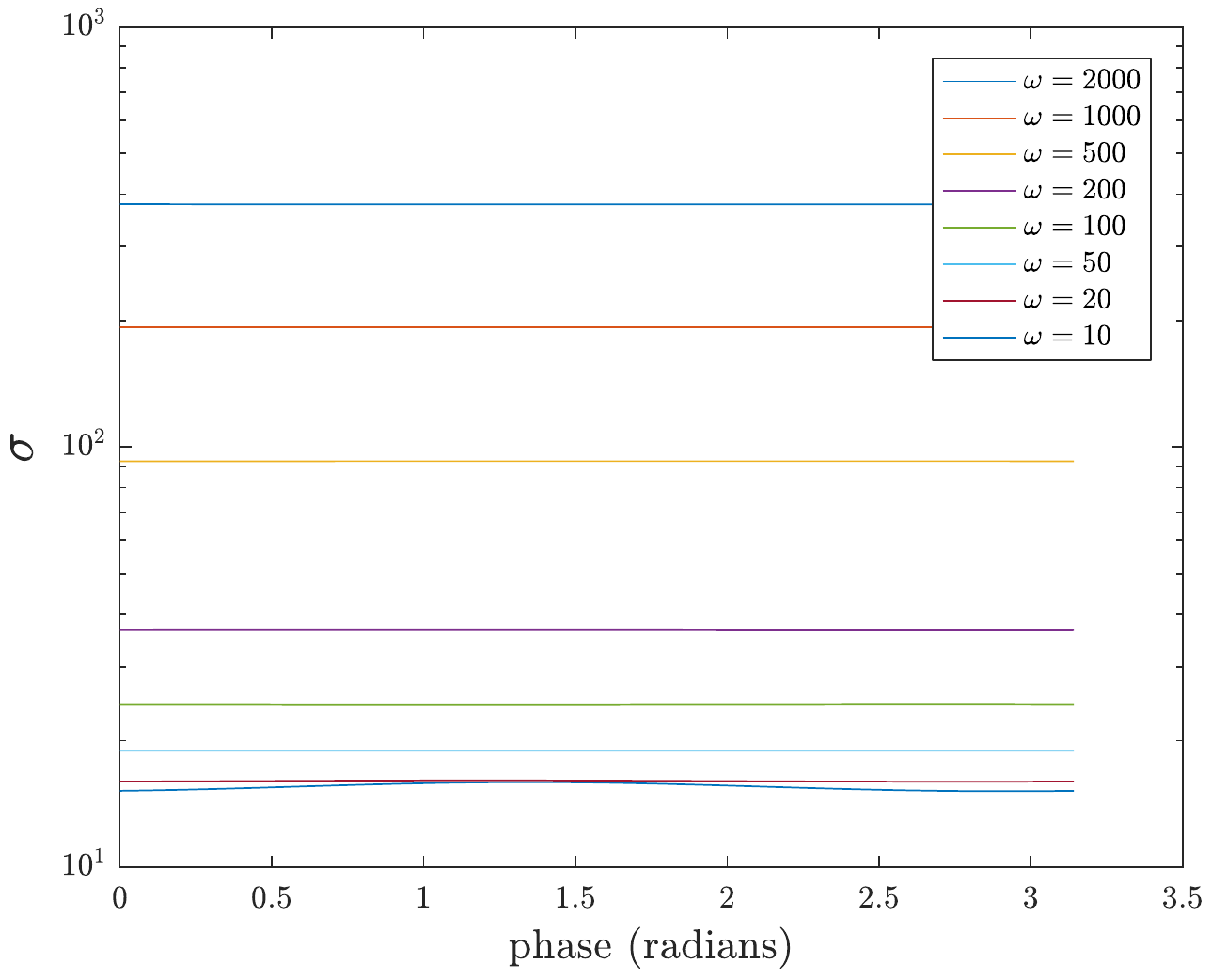}
	\caption{Forecast error bars $\sigma^{T+E}$ versus the phase $\phi$. Apart from the smallest frequency $\omega=10$, the error bar remains almost constant. This implies that the sine ($\phi=0$) and cosine ($\phi=\pi/2$) feature models can be constrained independently.}
	\label{forecast_phase_dependence}
\end{figure}

Figure \ref{forecast_phase_dependence} shows forecast error bars for the full phase range $[0,\pi]$ in the most sensitive experiment specification of 1' beam and 1$\mu K \cdot arcmin$ noise. The forecast $\sigma$ varies within 1\% level for every $\omega\ge20$. In terms of the Fisher matrix, the cross term $F_{sc}$ was 2-3 orders of magnitude smaller than $F_{ss}$ and $F_{cc}$ for all cases. In other words, correlation between the sine and cosine templates was smaller than 1\%. This justifies our previous choice of constraining $f_{NL}^{\sin}$ and $f_{NL}^{\cos}$ separately. We now focus our attention to $\sigma_{\sin}$ in future discussions.

For smaller values of $\omega$, the phase affects the error bar primarily through modulating the amplitude of the acoustic oscillations in the CMB itself. The radiation transfer functions are non-zero for $k$ values in $0 - 0.8 ~\text{Mpc}^{-1}$. The argument $\omega k$ covers less than two full periods in this $k$ range if $\omega \le 10\; \text{Mpc}$, and phase has a direct influence on the amplitude of the acoustic peaks. In the extreme case of $\omega=0$, bispectrum vanishes completely for the $\sin$ feature model. Variations in the overall bispectrum amplitude therefore result in varying Fisher information for low frequencies.

\subsection*{$l_{max}$ dependence}

\begin{figure}[ht]
	\centering
	\includegraphics[width=0.5\textwidth]{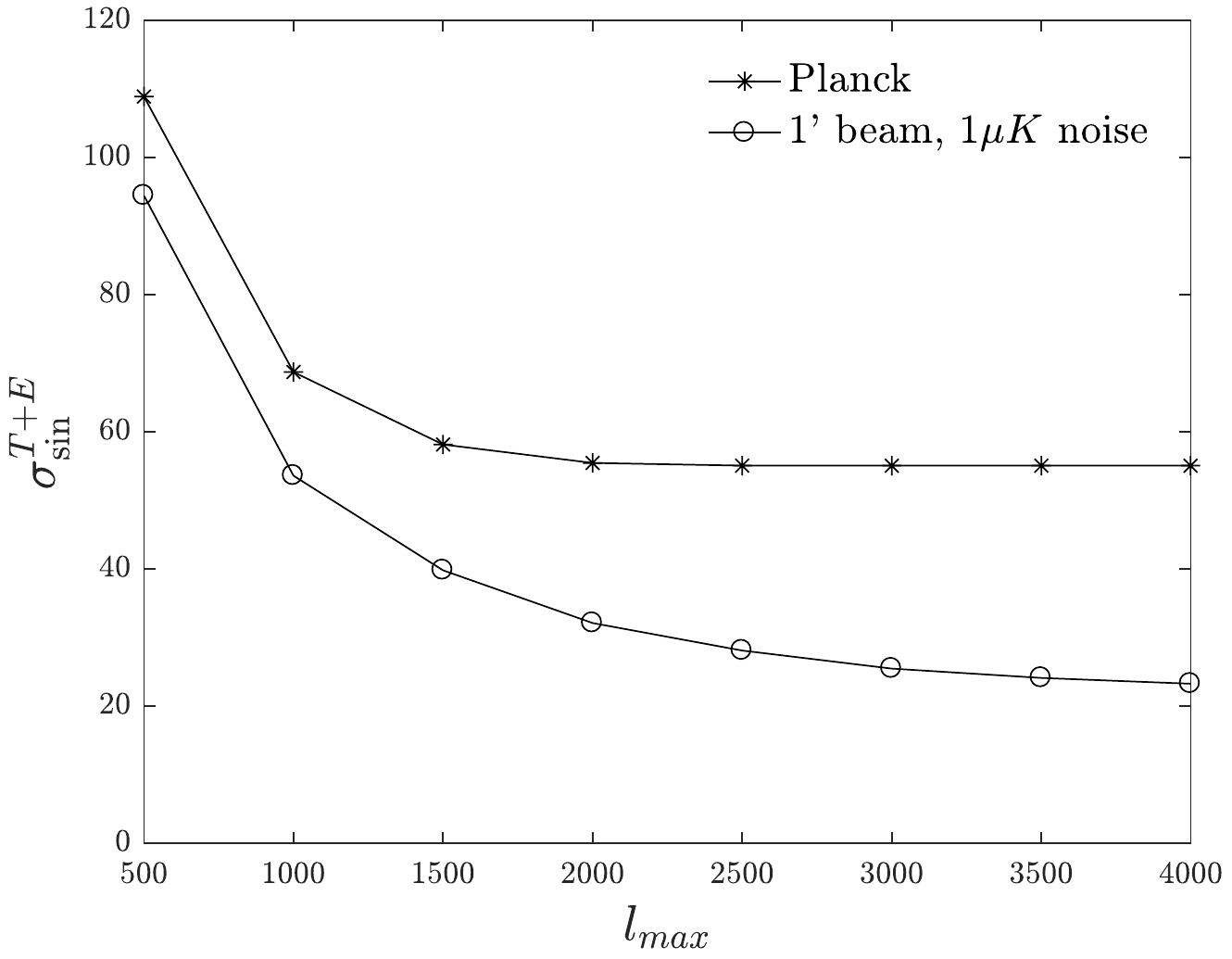}
	\caption{Forecast error bars $\sigma_{\sin}^{T+E}$ when multipoles $2\le l \le l_{max}$ are included, in comparison with Planck. The oscillation frequency $\omega$ is set to 100 Mpc in all cases. Planck did not have access to the information from modes $l\ge 2000$ due to large noise, but the CMB-S4 experiments are expected to be able to explore modes up to $l=4000$.}
	\label{forecast lmax dependence}
\end{figure}

Figure \ref{forecast lmax dependence} shows the graph of forecast error bar $\sigma_{\sin}^{T+E}$ as we increase $l_{max}$. The forecasts were done within angular scale range $2\le l \le l_{max}$, the oscillation frequency $\omega$ set to 100, and assuming 1' beam and 1$\mu K\cdot$arcmin noise. The Planck noise curves were approximated by ones for 5' beam and 47 $\mu K\cdot$arcmin noise for this plot only, since we extend $l_{max}$ to 4000 here. 

The Planck error bar essentially stalls out when $l_{max}$ reaches 2000. The forecast error bar, on the other hand, keeps decreasing until $l_{max}=4000$ thanks to the improved sensitivity in measuring small scale, or large $l$'s. Despite the information loss due to smaller sky coverage $f_{sky}$, the forecast error bar reduces to about 42\% of Planck by $l_{max}=4000$. This corresponds to a factor of 2.4 times improvement to measurement precision on $f_\text{NL}$.

\subsection*{Beam and noise dependence}

\begin{figure}[ht]
	\includegraphics[width=0.45\textwidth]{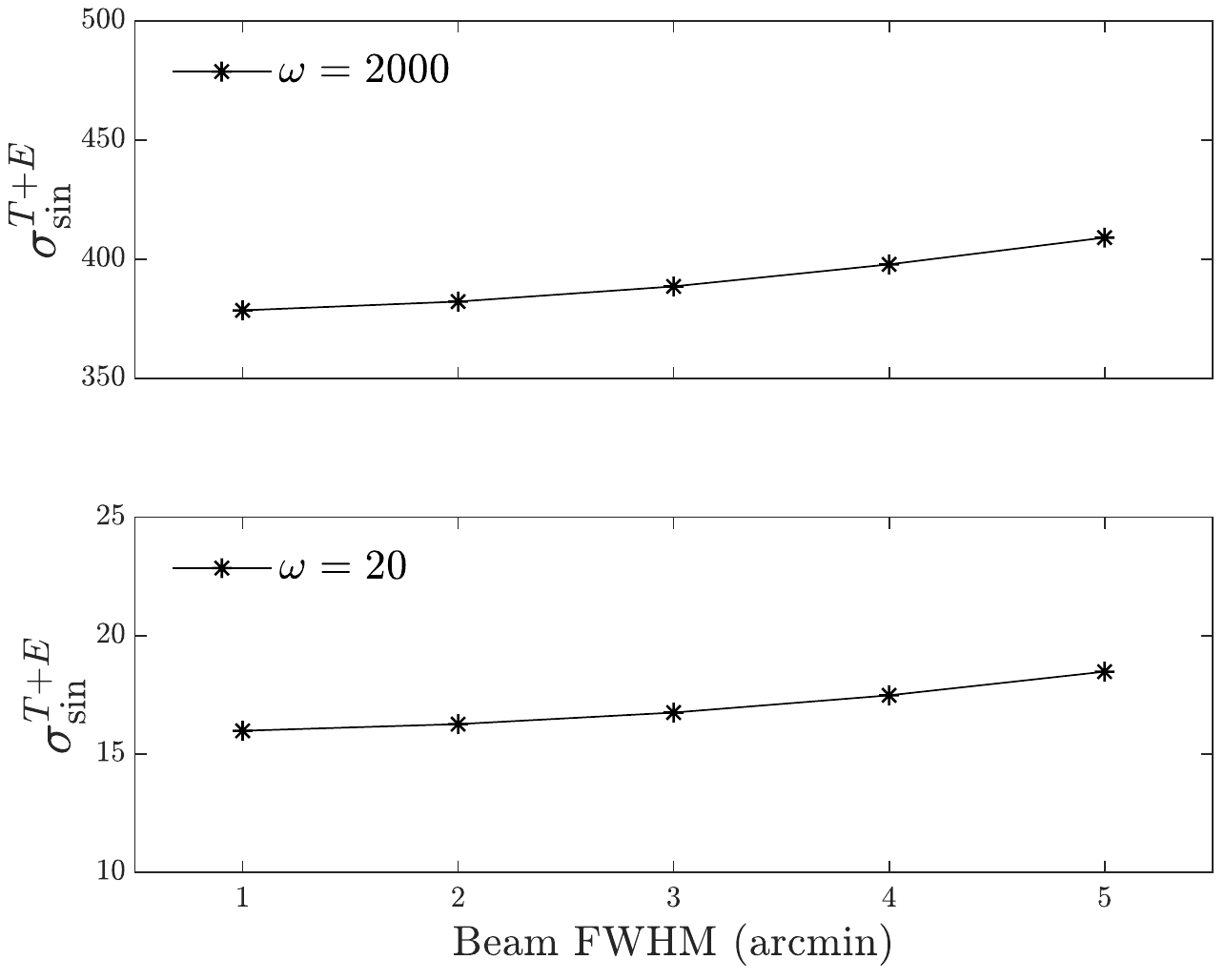}
	\includegraphics[width=0.45\textwidth]{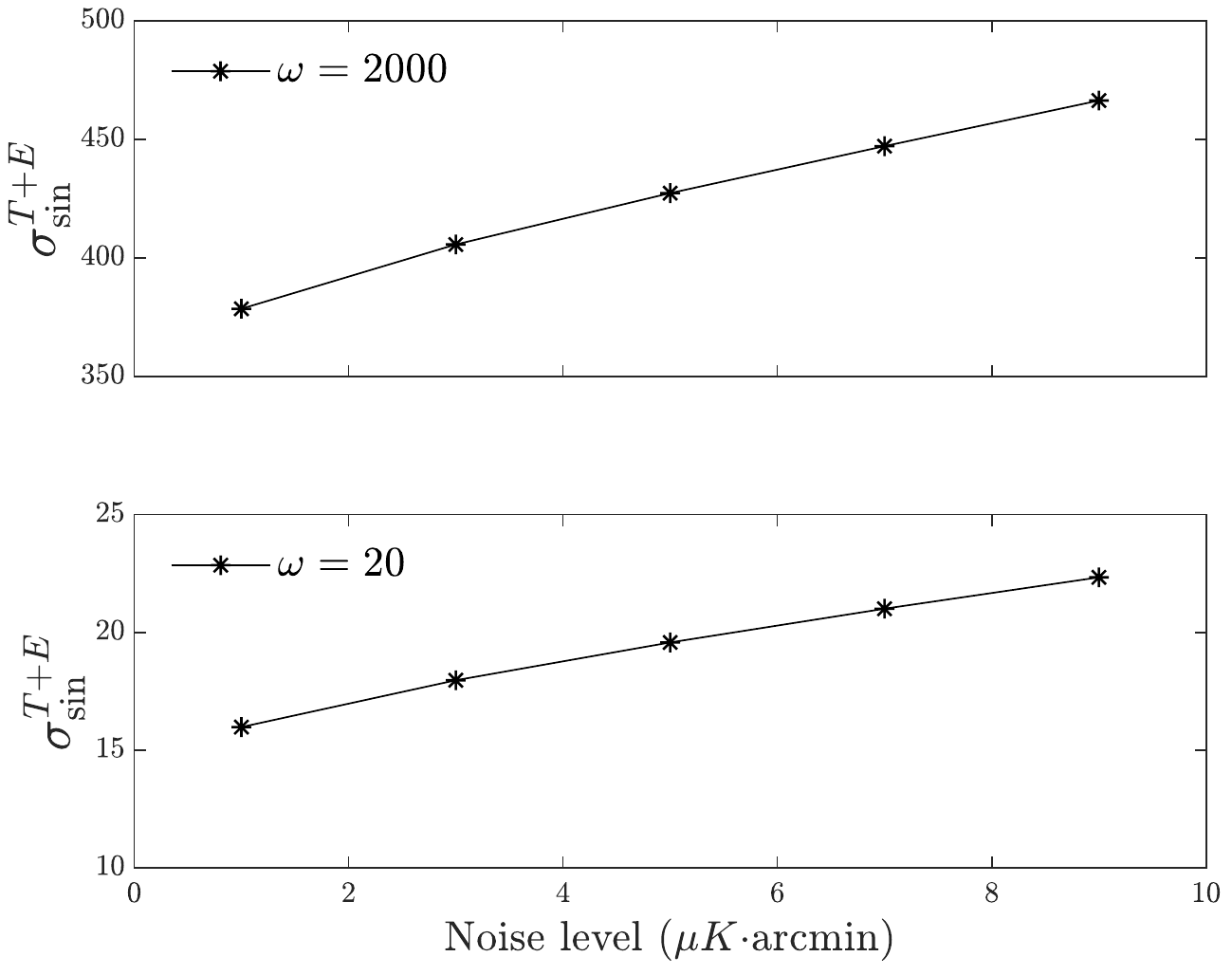}
	\caption{Beam (left) and noise (right) dependences of the forecast error $\sigma_{\sin}^{T+E}$ for fixed $\omega$. The noise level was set as 1$\mu K \cdot$arcmin for the first plot, while the second plot had fixed beam FWHM of 1'. We obtain less information from using wider beam and noisier sensors, as expected.}
	\label{forecast beam and noise dependence}
\end{figure}

We explore how different beam widths and noise levels affect the forecast error bars in this section. Figure \ref{forecast beam and noise dependence} shows forecast $\sigma_{\sin}^{T+E}$ for ranges of beam and noise levels. Their oscillation frequencies are also varied, but only two representatives $\omega=20$ and $2000$ are chosen here. Forecasts for the other values of $\omega$ also show similar dependences on beam width and noise level.

First of all, note that all estimated error bars in the plot are smaller than Planck, for which $\sigma_{\sin}^{T+E}=34$ when $\omega=20$ and $\sigma_{\sin}^{T+E}=610$ when $\omega=2000$. In fact even the least sensitive CMB-S4 specification of 5' beam and 9$\mu K\cdot$arcmin noise is expected to put better bounds on feature models.

Wider beams and noisier detectors provide less signal and thus larger error bars, as expected. In this range of beam width and noise levels, noise has a bigger effect on the forecast; experiments with 1' beam and 5$\mu K\cdot$arcmin noise yields larger error bars than the ones with 5' beam with 1$\mu K\cdot$arcmin noise. Between the most sensitive specification of 1' beam and 1$\mu K\cdot$arcmin and the least sensitive one with 5' beam and 9$\mu K\cdot$arcmin, $\sigma_{\sin}$ differs by a factor of 1.6.

\subsection*{Oscillation frequency dependence}

\begin{figure}[ht]
	\includegraphics[width=0.45\textwidth]{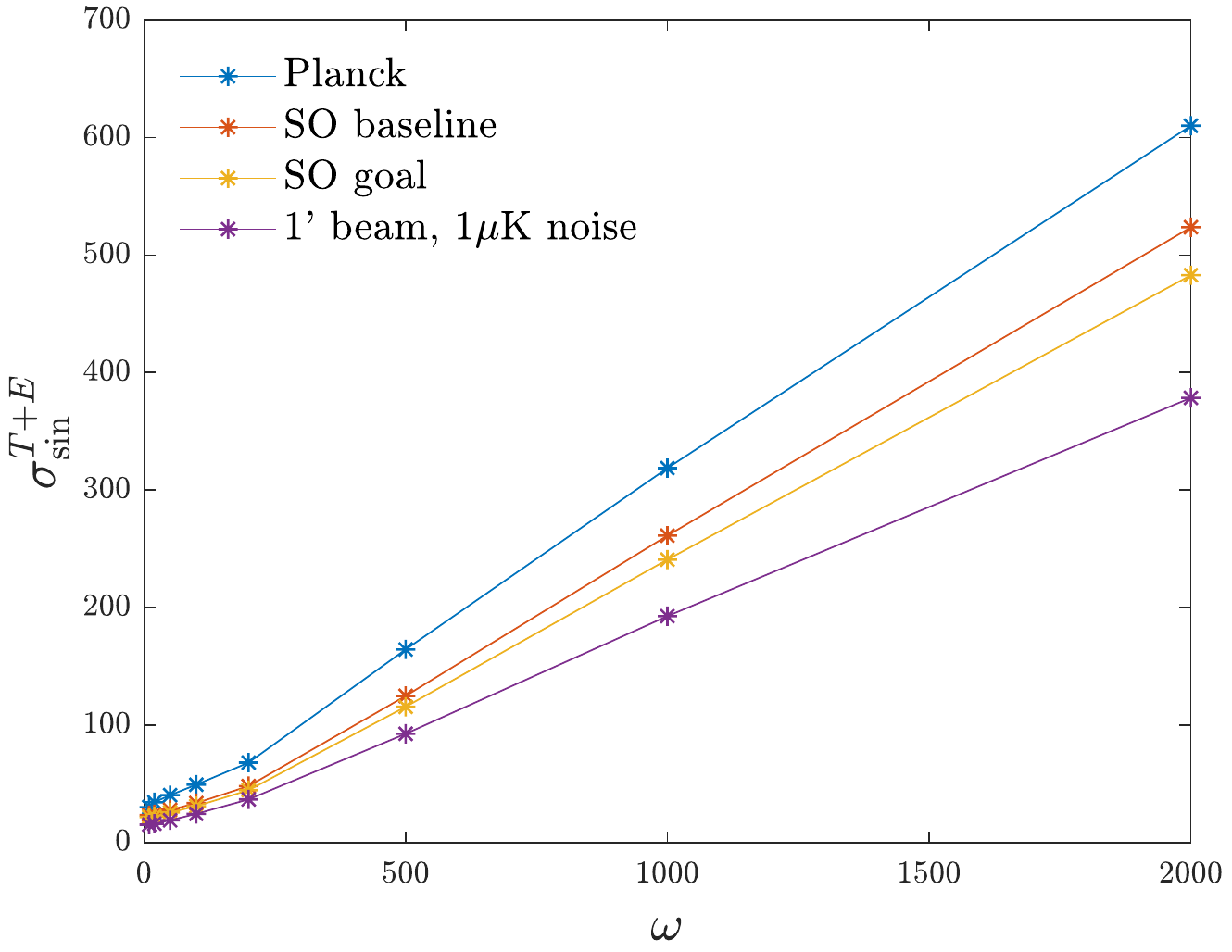}
	\includegraphics[width=0.45\textwidth]{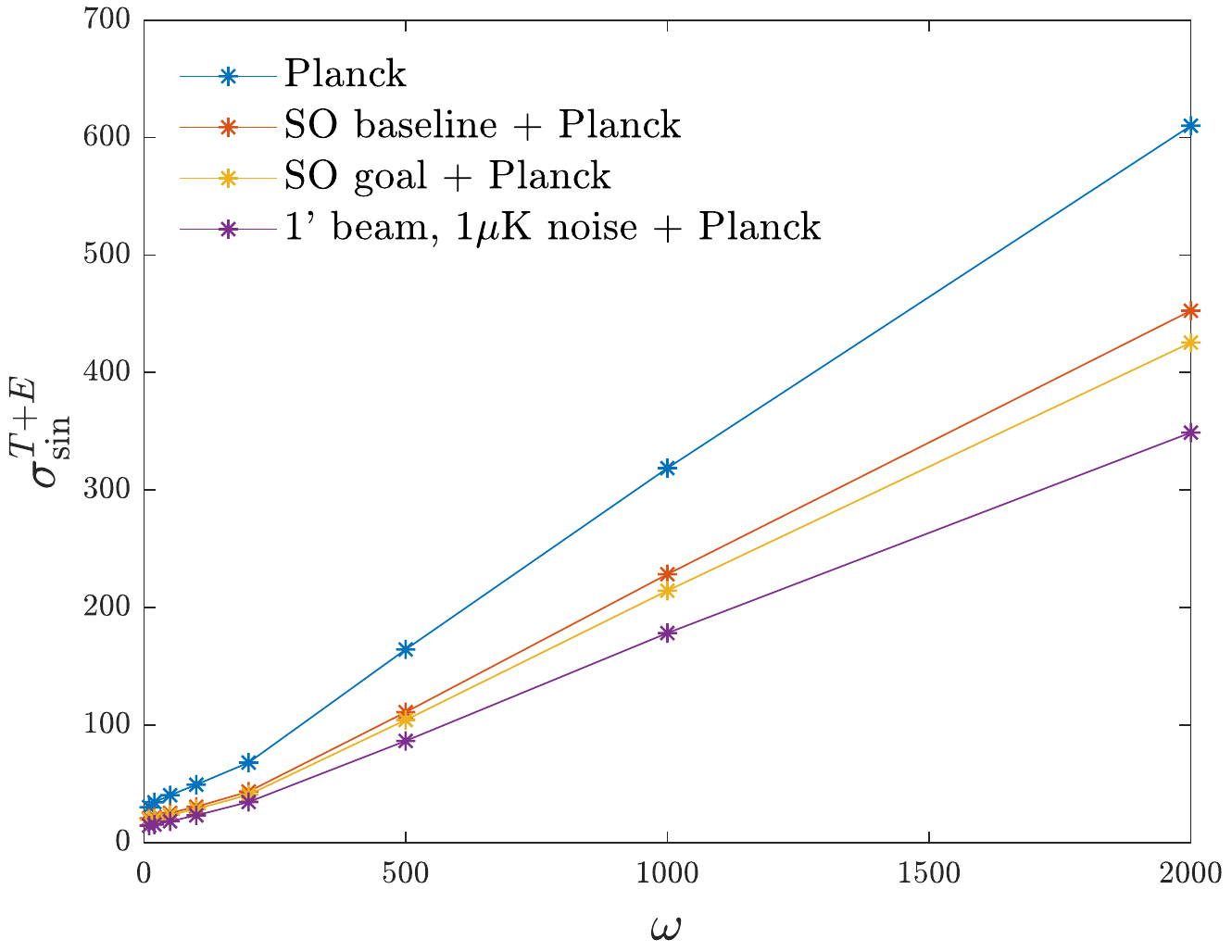}
	\caption{Frequency dependence of the forecast error in comparison to Planck (left). All CMB-S4 specifications would improve constraints on feature models. The most sensitive setup with 1' beam and 1$\mu K \cdot$arcmin noise is expected to yield error bars that are 1.6-2.1 times smaller than Planck. We get stronger constraints when the Planck results are combined with CMB-S4 (right).}
	\label{forecast omega dependence pol}
\end{figure}

We now present main results of the forecast. Figure \ref{forecast omega dependence pol} summarises the $\sigma_{\sin}$ forecasts for several different CMB-S4 preliminary specifications, including the Simons Observatory (SO) baseline and goal. Note that the 1/f noise effects are incorporated in SO forecasts but not in other ones. We also provide 1$\sigma$ errors for joint estimators, for which Planck signals from the fraction of the sky not covered by CMB-S4 are combined via $\sigma_\text{joint}^{-2} = \sigma_\text{CMB-S4}^{-2} +  \sigma_\text{Planck}^{-2}$. This method is not statistically optimal but sufficient to give an idea of the joint estimation power.

The most sensitive setup with 1' beam and 1$\mu K \cdot$arcmin noise would yield error bars that are 47-62\% of Planck, depending on the oscillation frequency $\omega$. These correspond to a factor of 1.6-2.1 improvement. Relatively smaller improvements are made for high oscillation frequencies. They correspond to smaller momentum scales $k_\ast = 2\pi/3\omega$, or larger angular scales, which benefit less from the increased sensitivity of CMB-S4 experiments. When the results are combined with Planck the error bar further reduces to 45-57\% of Planck, which is a factor of 1.7-2.2 improvement.

Forecast error bars from the SO baseline specification and the more ambitious one do not differ very much. Quoting in terms of the baseline values, $\sigma_{\sin}$ lies about 68-86\% of that of Planck or equivalently, 1.2-1.5 times smaller than Planck. Numbers change to 62-74\% when combined with Planck, so that the overall improvement ratio is about 1.3-1.6.

\begin{figure}[ht]
	\includegraphics[width=0.45\textwidth]{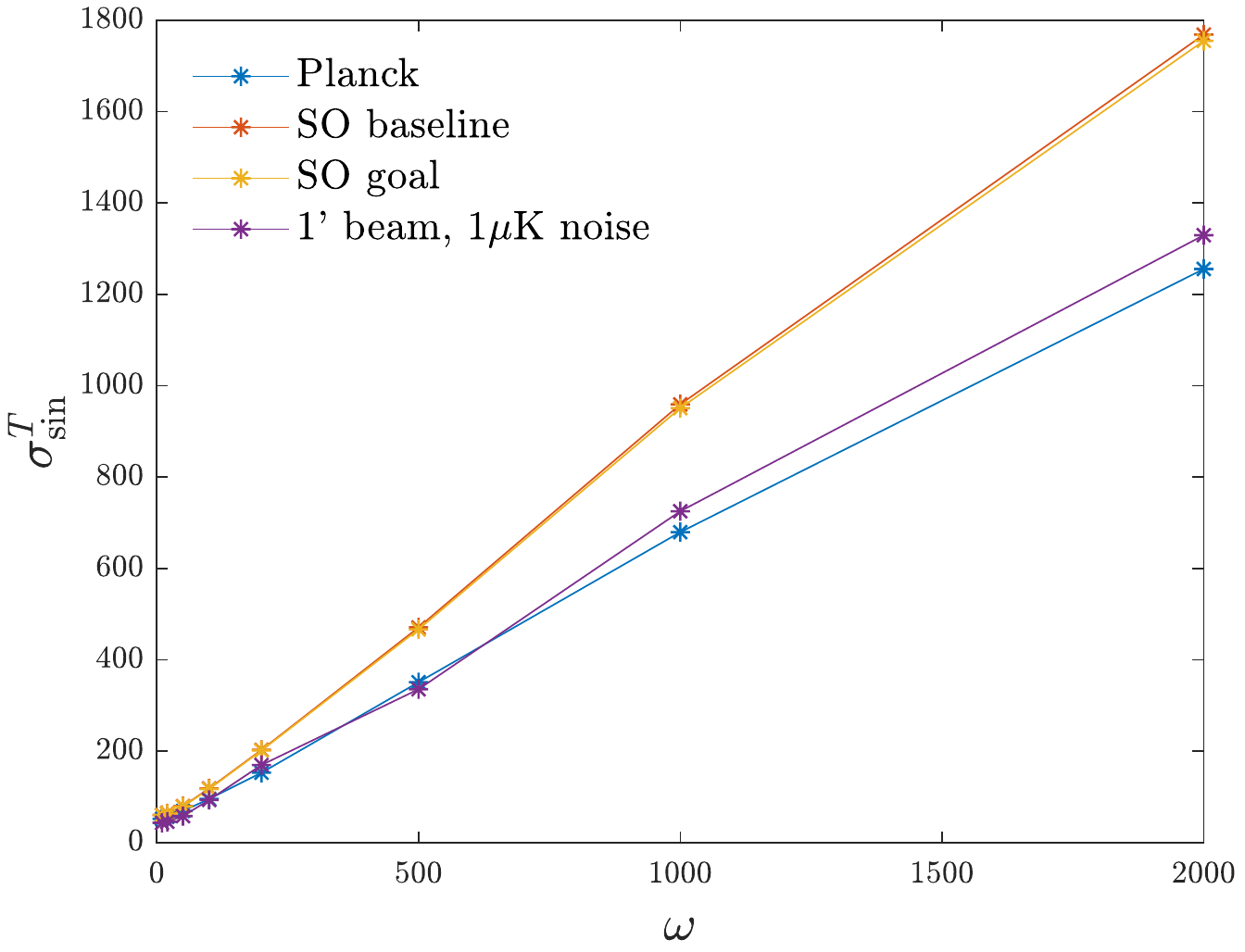}
	\includegraphics[width=0.45\textwidth]{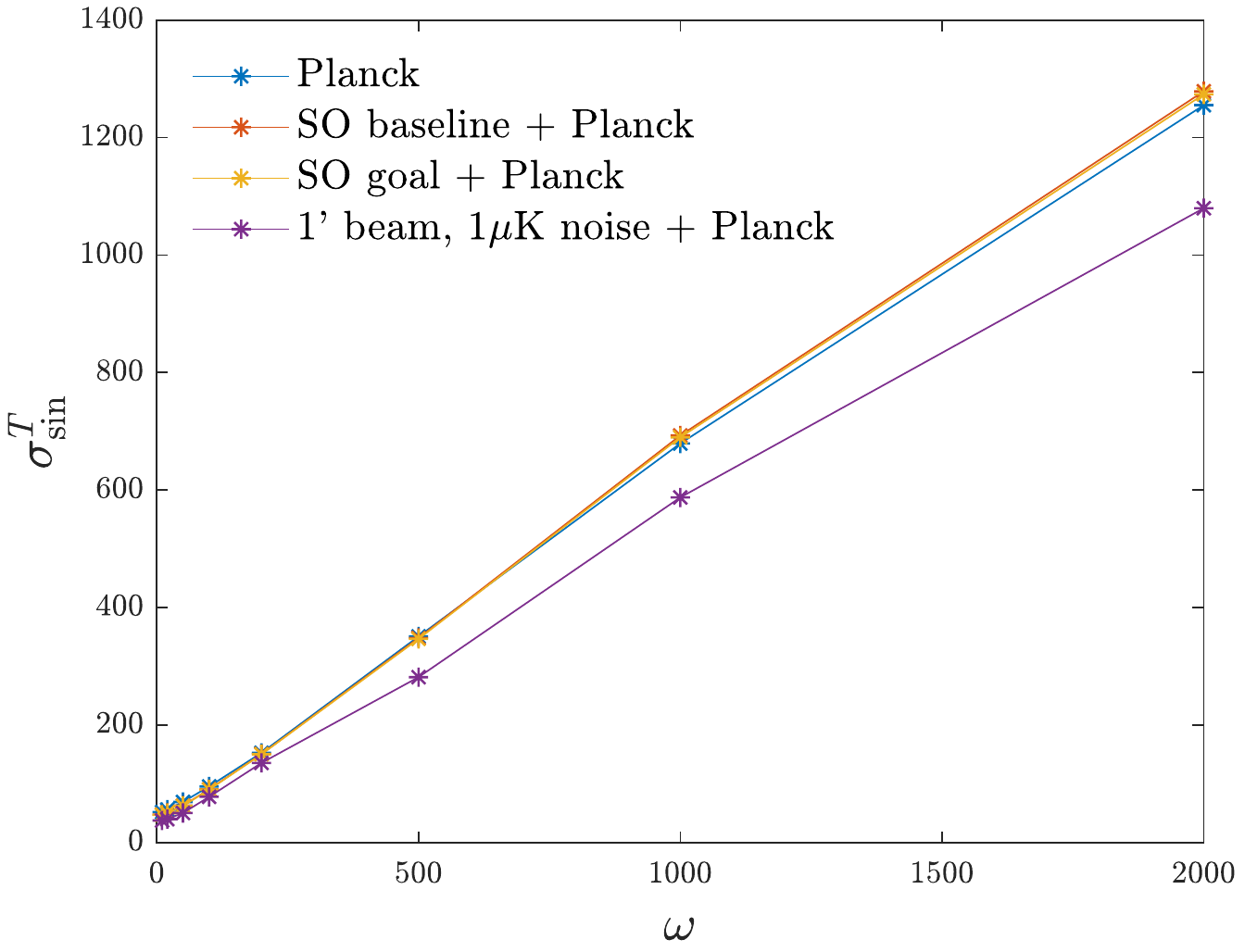}
	\caption{Frequency dependence of the forecast error from temperature data only, in comparison to Planck (left). The CMB-S4 experiments would perform worse than Planck when only the temperature map is concerned. After the addition of Planck data the error bars improve only marginally (right). Polarisation data are crucial in constraining feature models.}
	\label{forecast omega dependence T}
\end{figure}

Figure \ref{forecast omega dependence T} shows the results when only the CMB temperature data are used in the forecast. CMB-S4 would in fact be worse than Planck in terms of constraining $f_{NL}^\text{feat}$ for this case. The loss in information due to less sky coverage overwhelms the increased sensitivity. We see again that the real strength of CMB-S4 experiments lies in measuring the CMB polarisation.

\begin{figure}[ht]
	\includegraphics[width=0.45\textwidth]{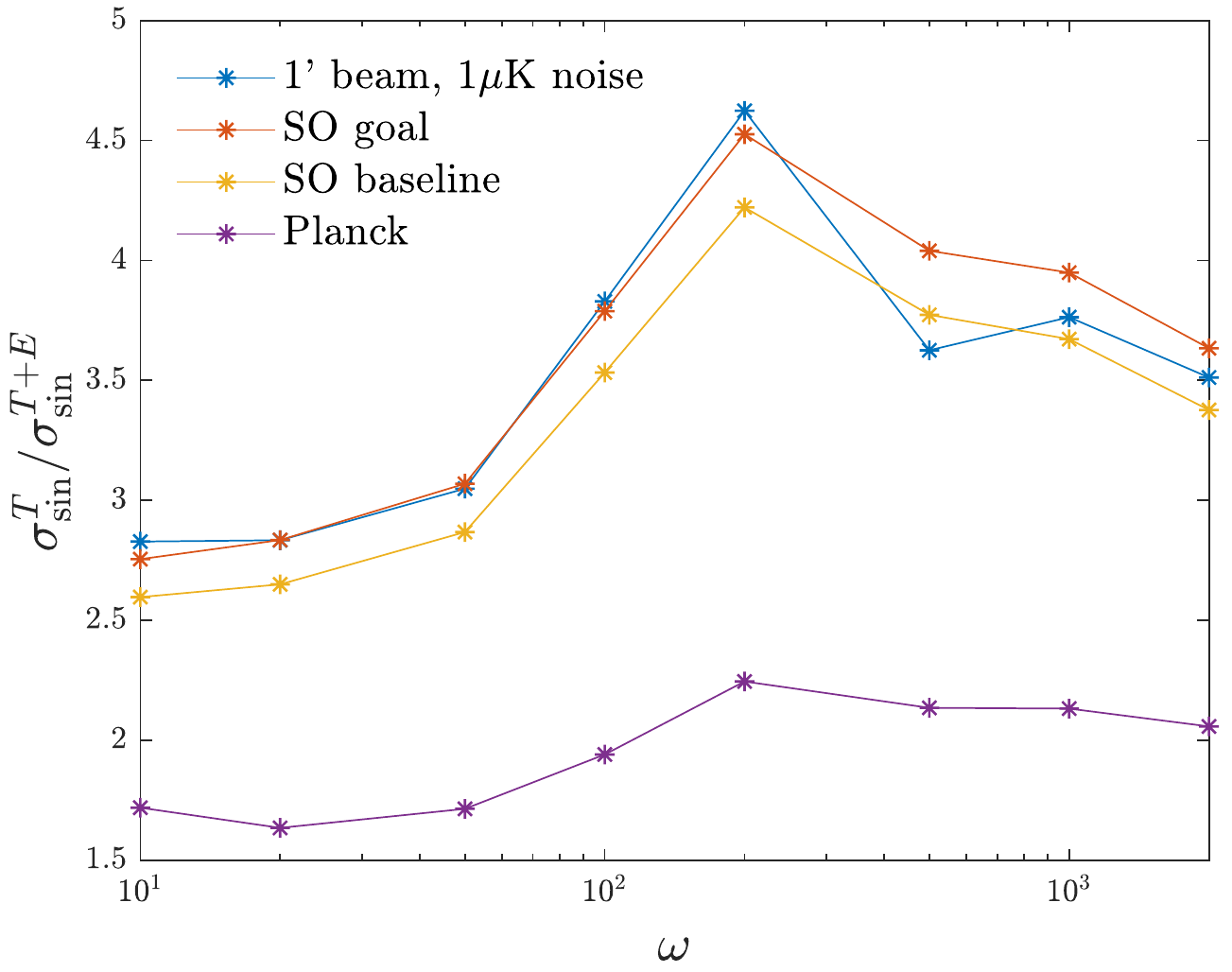}
	\includegraphics[width=0.45\textwidth]{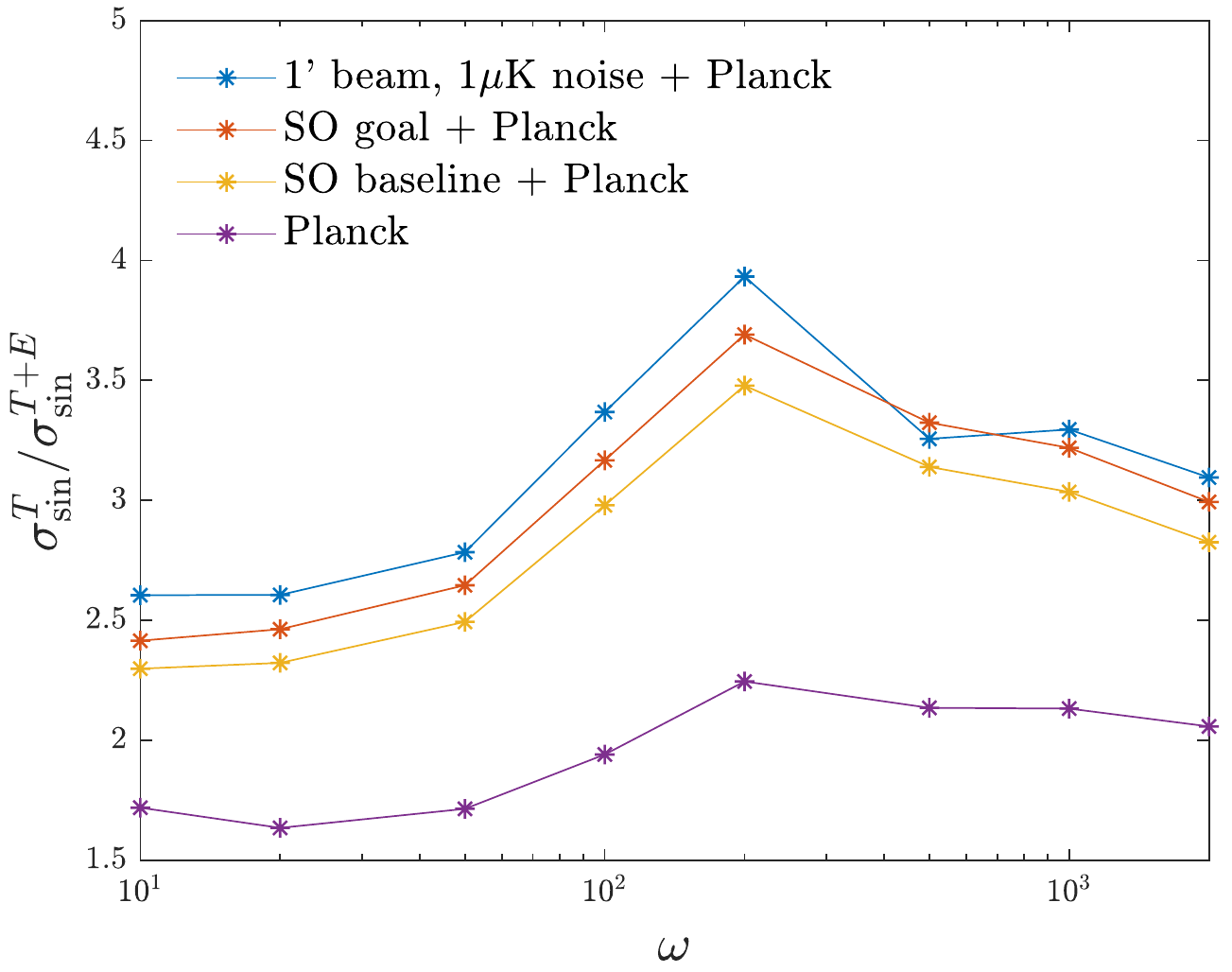}
	\caption{Improvements on the forecast error when including E-mode polarisation data. Constraints from the CMB-S4 experiments would improve significantly from addition of the polarisation data. The improvement is maximised around $\omega\approx200$ Mpc.}
	\label{forecast improvement ratio}
\end{figure}

Then how much information do we actually gain from adding E-mode polarisation? Figure \ref{forecast improvement ratio} shows the ratio of $\sigma_{\sin}$'s between the temperature-only (T) and polarisation-included (T+E) analyses. The forecast error bars reduces up to 4.6 times smaller when the polarisation information is added, which is much larger than the corresponding Planck value of 2.2. The ratio decreases overall when the joint statistics with Planck are considered. An intriguing feature of this plot is that the ratio is maximised around $\omega=200$ before it starts dropping again.

In order to gain insight on this behaviour, we performed some simplified computations using the power spectrum. We imposed oscillations on the primordial power spectrum as $P'(k) := P(k)(1+\sin(2\omega k + \phi))$, which is just like our feature model bispectrum template but with $\omega(k_1+k_2+k_3)$ replaced by $\omega(k+k)$. $P'(k)$ is then projected to the late-time harmonic space using the transfer functions;
\begin{equation}
	C_l'^{X_1 X_2} = \frac{2}{\pi} \int k^2 dk P'(k) \Delta_l^{X_1}(k) \Delta_l^{X_2}(k).
	\label{primordial power spectrum to cls}
\end{equation}
We observed that the fractional variation $(C_l'-C_l)/C_l$ displays some oscillations in $l$, and the largest contribution comes from a term $\propto \sin(2\omega l/\Delta\tau)$ where $\Delta\tau$ represents the conformal distance to last scattering surface. This fact can be explained by approximating the transfer function as $\Delta_l(k)\approx (1/3)j_l(k\Delta\tau)$ and noting that the spherical Bessel function has a sharp peak at $l$ for large $l$'s. The integral in (\ref{primordial power spectrum to cls}) therefore picks up a term proportional to $\sin(2\omega l/\Delta\tau)$.

\begin{figure}[ht]
	\centering
	\includegraphics[width=0.7\textwidth]{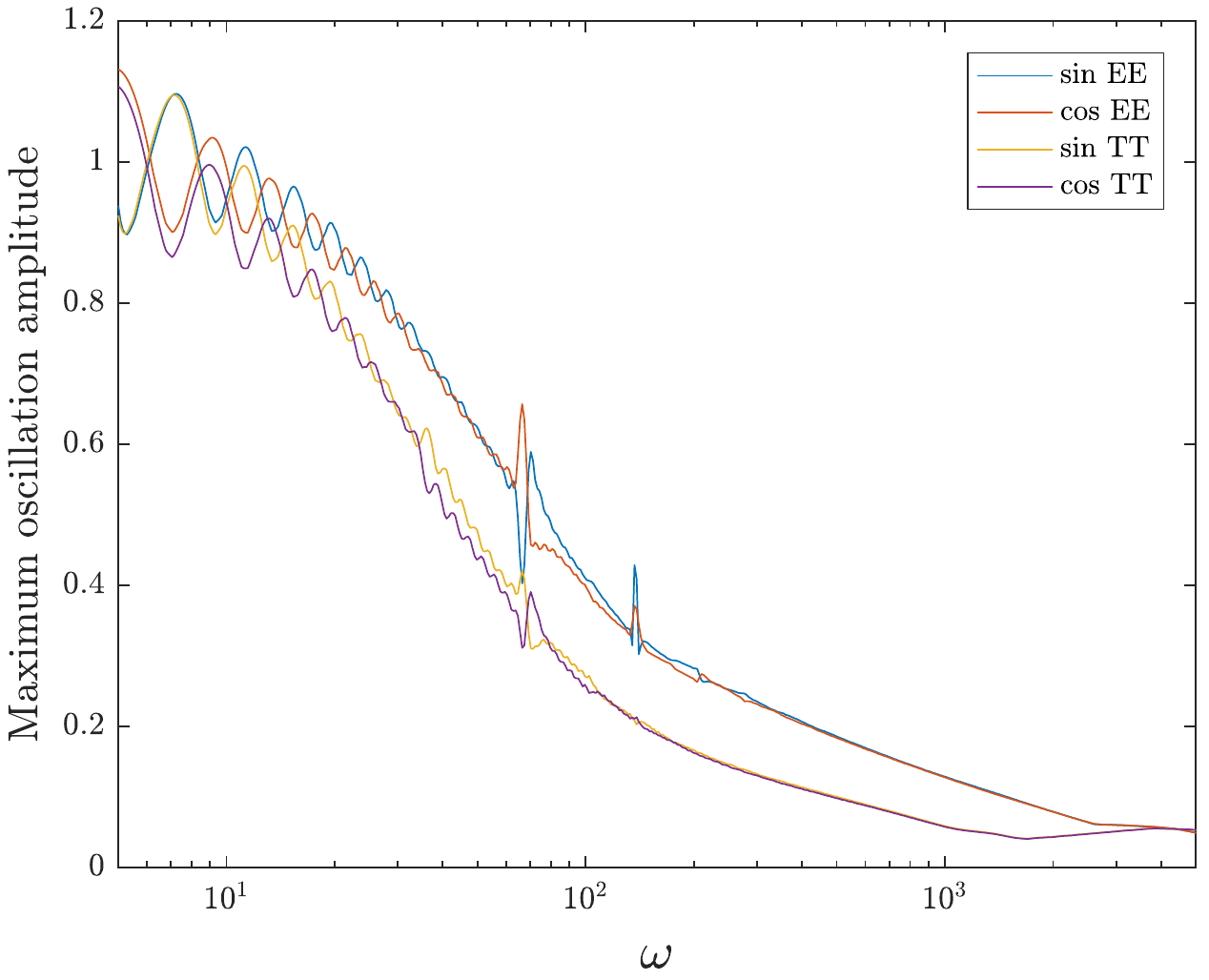}
	\caption{The maximum amplitude of oscillations detected in fractional variations of the projected power spectrum $C_l^{TT}$ and $C_l^{EE}$, when extra oscillations $\sin(2\omega k)$ and $\cos(2\omega k)$ were imposed on the primordial power spectrum. Heuristically this shows that the E-mode polarisation is more sensitive to the primordial oscillations, especially in the $\omega$ range of 70 to 300.}
	\label{insight feature plot}
\end{figure}

The amplitude of these `maximal' oscillations in $(C_l'-C_l)/C_l$ were then computed using discrete Fourier transform for different values of oscillation scale $\omega$ and two different phases $\phi=0,\pi/2$ (i.e. sine and cosine). The results are shown in Figure \ref{insight feature plot}. Some extra wiggles to the graph come from the phase of oscillations imposed; we indeed see that graphs of sine and cosine oscillate between each other. Some peak features near $\omega\approx70$ and 140 arise from resonances with Baryonic Acoustic Oscillations.

We can think of the computed amplitude as a measure of information $C_l$'s contain about primordial oscillations. First of all, note that the amplitude in all four plots generally decreases as $\omega$ grows. Previously in Figure \ref{forecast omega dependence pol} we saw that the amount of information obtained from the CMB is smaller for larger $\omega$'s, consistent with what can be said from the amplitude analysis. Moreover, the amplitudes for the EE mode are generally larger than the TT mode ones, and their difference is the largest in the $\omega$ range of 70 to 300. This could serve as a heuristic explanation for the improvement in forecast error bars from including polarisation data being maximised around $\omega=200$, as depicted in Figure \ref{forecast improvement ratio}.

\subsection*{Comparison to scale invariant models}

Our pipeline for forecasting $f_{NL}^\text{feat}$ also yields forecasts for $f_{NL}$ of the constant model. Constant models are scale invariant and have a trivial shape, so that $B(k_1,k_2,k_3) \propto (k_1 k_2 k_3)^{-2}$.
Forecasts on $f_{NL}^\text{const}$ follow from out pipeline by simply setting oscillation frequency$\omega = 0$ and phase $\phi = \pi/2$. Table \ref{forecast constant model} summarises the forecast results for several different CMB-S4 specifications mentioned before, using both T and E data and in combination with Planck data from the regions of the sky not covered by CMB-S4. For the 1' beam and 1$\mu K \cdot$arcmin noise setup, the error bar is expected to be reduced by a factor of 2.3 compared to Planck.

\begin{table}[ht]
	\centering
	\renewcommand{\arraystretch}{1.4}
	\begin{ruledtabular}
		\begin{tabular}{ccccc}
			& Planck  & SO baseline + Planck & SO goal + Planck & 1' beam, 1$\mu K$ noise + Planck \\ \hline
			$\sigma(f_\text{NL}^\text{const})$ & 23.4 & 14.9 & 14.0 & 10.4            \\ 
		\end{tabular}
	\end{ruledtabular}
	\caption{Forecasts on the estimation errors of $f_{NL}$ for the constant model}
	\label{forecast constant model}
\end{table}

The latest Planck constraints on $f_\text{NL}$ of some popular bispectrum templates are given by $f_\text{NL}^\text{local} = 2.5 \pm 5.7$, $f_\text{NL}^\text{equil} = -16 \pm 70$, and $f_\text{NL}^\text{ortho} = -34 \pm 33$ \cite{PlanckCollaboration2015}. CMB-S4 experiments are expected to yield better estimates on these as well. Table \ref{forecast various models} summarises the forecast improvement ratio given in \cite{Abazajian2016} together with the constant and feature model ratios computed in this work.

\begin{table}[ht]
	\centering
	\renewcommand{\arraystretch}{1.4}
	\begin{ruledtabular}
		\begin{tabular}{cccccc}
			& Local & Equilateral & Orthogonal & Constant & Feature ($\omega=200$) \\ \hline
			$\sigma^\text{Planck}/\sigma^\text{CMB-S4}$ & 2.5   & 2.1         & 2.4        & 2.3      & 2.0                
		\end{tabular}
	\end{ruledtabular}
	\caption{Expected improvements on estimation errors of $f_\text{NL}$ for the CMB-S4 1'beam, 1$\mu K\cdot$arcmin setup, for various bispectrum templates. The local, equilateral and orthogonal results are quoted from \cite{Abazajian2016}.}
	\label{forecast various models}
\end{table}

To the authors' surprise, the estimation error for feature models does not improve as much as other templates. Feature models benefit much more from polarisation data than other scale independent shapes; for example, $\sigma^{T}/\sigma^{T+E} = 4.6$ for the feature model with $\omega=200$ in CMB-S4, while the value equals 2.8 for the constant model. Because CMB-S4 would have significantly enhanced polarisation measurement sensitivity, we originally expected the feature models to be constrained significantly better than Planck.

In order to investigate this lack of improvement, we performed a breakdown analysis on the improvements gained from CMB-S4 temperature and polarisation; we computed $\sigma(f_\text{NL})$ for the constant and feature models using each of the four combinations of Planck / CMB-S4 noise curves for temperature / polarisation (e.g. Planck T + CMB-S4 E). The results are summarised in Table \ref{forecast mixed}.

\begin{table}[ht]
	\centering
	\renewcommand{\arraystretch}{1.4}
	\parbox{.45\linewidth}{
		\begin{ruledtabular}
			\begin{tabular}{cccc}
				\multicolumn{2}{c}{\multirow{2}{*}{$\sigma(f_\text{NL}^\text{const})$ improvement}}  &  \multicolumn{2}{c}{E} \\ 
				\multicolumn{2}{c}{} & Planck & CMB-S4 \\ \hline
				\multirow{2}{*}{T} & Planck & 1.0 & 1.6 \\
				& CMB-S4 & 1.1 & 2.2
			\end{tabular}
		\end{ruledtabular}
	}
	\parbox{.45\linewidth}{
		\begin{ruledtabular}
			\begin{tabular}{cccc}
				\multicolumn{2}{c}{\multirow{2}{*}{$\sigma(f_\text{NL}^\text{feat})$ improvement}} &  \multicolumn{2}{c}{E} \\ 	\multicolumn{2}{c}{} & Planck & CMB-S4 \\ \hline
				\multirow{2}{*}{T} & Planck & 1.0 & 1.7 \\
				& CMB-S4 & 0.9 & 1.9
			\end{tabular}
		\end{ruledtabular}
	}
	\caption{Expected improvements on the estimation errors of $f_\text{NL}$ for each combination of Planck / CMB-S4 temperature (T) and polarisation (E) data. Here the CMB-S4 assumes 1' beam and 1$\mu K$arcmin noise. For feature model the oscillation frequency $\omega=200$ and phase $\phi=0$. The sky fraction $f_{sky}=0.4$ for all cases except for Planck T + Planck E.}
	\label{forecast mixed}
\end{table}

We see that the constraints on feature models improve by a factor of 1.7 when swapping Planck polarisation noises with the CMB-S4 ones. This factor is indeed larger than that of the constant model, which equals 1.6. The difference is however not significant. It seems that the amount of feature signals in polarisation data left unexplored by Planck is not tremendously large compared to the constant model. The feature model improves less than the constant model when the temperature measurements are enhanced. In fact, for feature models the signal loss from smaller sky fraction $f_{sky}$ eclipses the signal gain from more sensitive temperature measurements. This lack of improvements from temperature causes the full CMB-S4 constraints on the feature model not to improve as much as the constant model overall.

\section{Conclusion} \label{section: conclusion}

Upcoming CMB Stage-4 experiments will provide an opportunity to measure CMB temperature and polarisation with greater precision. The estimation of primordial non-Gaussianity parameters would greatly benefit from the improvement in measurement sensitivity. In this research we made forecasts on $f_\text{NL}$ for the feature models, which have not been done so far despite the growing interests on inflation models with primordial oscillations. For efficient forecasts we simplified the bispectrum estimator for $f_\text{NL}$ by orthonormalising the covariance matrix, further optimising the computation. When the most sensitive CMB Stage-4 experiment specification of 1' beam and 1$\mu K$arcmin noise is concerned, we expect a factor of 1.7-2.2 times more stringent constraints compared to Planck. Under realistic Simons Observatory conditions the improvement would be about 1.3-1.6 to Planck.

Although this is not a massive boost in the estimation power, we can hope to verify current 4$\sigma$-level signals found in the 2015 Planck analysis. It is also worth noting that the CMB-S4 experiments would allow us to explore higher $l$ modes, especially since localised oscillations in this range are currently unconstrained. Moreover, though we have only considered linearly spaced oscillations in this work, we expect even better improvements on the models inducing log spaced oscillations. Higher $l$ modes would promote the constraining power as the oscillation slows down in small scales for this type of models. Lastly, cross-validation using these new statistically independent modes would be useful.

We also extensively studied how the forecasts depend on various parameters. Frequency dependences of the ratio between T and T+E forecasts were particularly illuminating - the improvement from adding polarisation information is maximised around $\omega = 200$. Some simplified calculations were presented to heuristically address this fact. Even though the estimation power on feature models massively benefit from the polarisation data, overall expected improvements compared to Planck are quite underwhelming. Breakdown analysis on temperature and polarisation contribution revealed that the feature models would indeed improve more than other scale-independent models if only the polarisation measurement sensitivity is enhanced to the CMB-S4 standards. However, boosts in the temperature measurements affect scale-independent models more so that they gain more information overall.

%\printbibliography
\bibliography{references}

\end{document}